\documentclass[letterpaper]{aa}
\usepackage{txfonts}

\usepackage{psfig}
\usepackage[dvips]{graphicx}
\usepackage[]{longtable}
\usepackage{natbib}
\bibpunct{(}{)}{;}{a}{}{,}

\def\lya{Ly$\alpha$}

\def\kms{~km~s$^{-1}$}

\def\al2{Al{\sc ii}$\lambda$167.0}
\def\si2{Si{\sc ii}$\lambda$152.6}
\def\mg2{Mg{\sc ii}$\lambda\lambda$279.6,280.3}

\def\thetae{\hbox{$\theta_{\mathrm{E}}$}}

\def\masse{\hbox{$M_{\mathrm{E}}$}}
\def\hubble{\hbox{$h_{70}$}}
\def\h70m1{\hbox{$h_{70}^{-1}$}}
\def\kmsmpc{\hbox{km s$^{-1}$ Mpc$^{-1}$}}
\def\hm2{\hbox{$h_{70}^{-2}$}}

\def\msun{\hbox{$M_{\odot}$}}
\def\mlbsun{\hbox{$M_{\odot}/L_{B\odot}$}}

\def\deg{\hbox{$^\circ$}}
\def\arcmin{\hbox{$^\prime$}}
\def\arcsec{\hbox{$^{\prime\prime}$}}
\def\utw{\smash{\rlap{\lower5pt\hbox{$\sim$}}}}
\def\udtw{\smash{\rlap{\lower6pt\hbox{$\approx$}}}}

\def\fh{\hbox{$.\!\!^{\rm h}$}}
\def\fm{\hbox{$.\!\!^{\rm m}$}}
\def\fs{\hbox{$.\!\!^{\rm s}$}}
\def\fdeg{\hbox{$.\!\!^\circ$}}

\def\farcsec{\hbox{$.\!\!^{\prime\prime}$}}

\begin{document}

\title{Discovery of a high-redshift Einstein ring\thanks{Based on 
observations carried out at the ESO/VLT 
under programs 68.A-01706(A), 71.A-0102(A) and 72.A-0111(A) 
and at La Silla 3.6-m telescope under DDT.}}

\author{
R. A. Cabanac\inst{1,2,3},
D. Valls-Gabaud\inst{3,4},
A. O.\ Jaunsen\inst{2},
C. Lidman\inst{2}, and
H. Jerjen\inst{5} }

%\offprints{cabanac@cfht.hawaii.edu}

\institute{Dep. de Astronom\'{\i}a y Astrof\'{\i}sica, %PUC, 
Pontificia Universidad Cat\'{o}lica de Chile, 
Casilla 306, Santiago, Chile
\and
European Southern Observatory, Alonso de Cordova, 3107, 
Casilla 19001, Santiago, Chile
\and
Canada-France-Hawaii Telescope, 65-1238 Mamalahoa Highway, 
Kamuela, HI 96743, USA
\and
CNRS UMR 5572, LATT, 
Observatoire Midi-Pyr\'en\'ees, 14 Av. E. Belin, 31400 Toulouse, France
\and
Research School of Astronomy and Astrophysics,  
ANU, Mt. Stromlo Observatory, 
Weston ACT 2611, Australia}

%-------------------------------------------------------------------------
\date{Received ..........  ; accepted .......... }
\titlerunning{A high-redshift Einstein ring}
\authorrunning{Cabanac et al.~}
%-------------------------------------------------------------------------

\abstract{We report the discovery of a partial Einstein 
ring of radius 1\farcsec48 produced by a massive (and seemingly isolated) 
elliptical galaxy. The spectroscopic follow-up at the VLT reveals a 
2$L_\star$ galaxy at $z=0.986$, which is lensing a post-starburst 
galaxy at $z=3.773$. This unique configuration yields a very precise measure 
of the mass of the lens within the Einstein radius, 
$(8.3 \pm 0.4) \, 10^{11}$\h70m1~\msun. The fundamental plane 
relation indicates an evolution rate of $d \log (M/L)_\mathrm{B} / dz = -0.57\pm0.04$,
similar to other massive ellipticals at this redshift. The source galaxy shows strong 
interstellar absorption lines indicative of large gas-phase metallicities, 
with fading stellar populations after a burst. 
Higher resolution spectra and imaging will allow the detailed
study of an unbiased representative of the galaxy population when the 
universe was just 12\% of its current age. 

\keywords{cosmology: observations -- gravitational lensing -- 
galaxies : high-redshift - ellipticals - evolution -- FOR J0332-3557}
} 

\maketitle

\section{Introduction}

One of the key issues in galaxy formation within the current
$\Lambda$CDM framework of structure formation is
the mass assembly histories of galactic halos. In this 
paradigm, the growth of halo mass through mergers produces giant
galaxies and star formation appears
rather late during this process. Measuring the evolution
of the mass-to-light ratio hence constrains directly this formation 
scenario, and provides clues on the evolution of the
fundamental plane. 
Various deep surveys have uncovered different galaxy populations, but the
selection criteria produce biased samples :
UV-selected \citep{steidel03} and narrow-band selected \citep{hu02} 
samples are sensitive to actively star-forming galaxies and 
biased against quiescent, evolved systems,
while sub-millimeter  \citep{blain00}
 and near-infrared surveys \citep{mccarthy01,abraham04}
 select dusty starburst galaxies and very red galaxies respectively.
It is not clear how representative
these samples are of the population of distant galaxies as a whole and how they
are related to present-day galaxies. In contrast, selection through 
gravitational lensing is not biased, as 
any distant galaxy which happens to have a massive deflector
along the line of sight can be amplified. 
This technique has yielded many examples of very distant galaxies, 
amplified by the mass distribution of foreground clusters of galaxies.
\citep{ellis01,pettini02,pello04}.
Configurations where lenses are isolated and distant ($z \sim 1$) 
galaxies are much rarer and yet provide direct measures
of the total mass without any  assumptions on stellar evolution. 
Whilst the total mass enclosed within the projected
Einstein radius \thetae~ can be well measured with systems presenting a few 
($\sim$2-4) images of the lensed galaxy, many more constraints are given by 
the observations of 
nearly-complete Einstein rings \citep{kochanek01}, as
 they do not suffer from the well-known ellipticity-shear degeneracy due to 
the many data along different position angles. %which provide
%more constraints on the higher-order moments of the mass distribution.
% (and not only the monopole).

Although many arcs have been discovered, associated with massive
galaxy clusters and their dominant central galaxies, there are very
few optical rings \citep{miraldaescude92} or arcs, especially at high 
redshifts. Only a few systems with lenses above $z_{\mathrm{l}} \sim 0.9$  are known : 
CFRS 03.1077 ($z_\mathrm{l}=0.938, z_\mathrm{s}=2.941$, \citealt{crampton02}),
MG 0414+0534 ($z_\mathrm{l}=0.958, z_\mathrm{s}=2.62$, \citealt{hewitt92}; \citealt{tonry99}),
MG 2016+112 ($z_\mathrm{l}=1.004, z_\mathrm{s}=3.273$, \citealt{lawrence84}; 
\citealt{koopmans02}),
and possibly J100424.9+122922 ($z_\mathrm{l}\sim 0.95, z_\mathrm{s}=2.65$, \citealt{lacy02})
and GDS J033206$-$274729 ($z_\mathrm{l}\sim $ 0.96, \citealt{fassnacht04}).
In all cases the arcs cover less than 60 degrees around the central lens and
hence are not nearly-complete Einstein rings. The only optical ring reported 
so far \citep{warren96}, over some 170\deg~ with a radius 
1\farcsec08, is a galaxy at $z_\mathrm{s}=3.595$  lensed by an elliptical galaxy at 
a much smaller $z_\mathrm{l}=0.485$. 
% A similar long arc of 120\deg~
% produced by a field elliptical at $z_\mathrm{l}=0.617$ falls in the same category 
% (Blaskeslee et al. \cite{blakeslee04}) of lower redshift lenses.

Here we report  the discovery of a fourth confirmed system, 
dubbed FOR J0332$-$3557 (03\fh32\fm59\fs94, $-$35\deg57\arcmin51\farcsec7, 
J2000), in a sight line through the outskirts of the Fornax cluster, where the 
reddening of our Galaxy is $E(B-V)=0.02$ \citep{schlegel98}.
This new system is remarkable on two accounts. First, it is a bright, 
almost complete Einstein ring of radius 1\farcsec48, 
covering some 260\deg~ around the lens, extending over 
 4\arcsec$\times$3\arcsec\, (\S2)
and with a total apparent magnitude $R_\mathrm{c}=22.2$. As discussed in \S3, the lens
appears to be an elliptical galaxy at $z_\mathrm{l}=0.986$.
Second, the lensed source is a galaxy at redshift 
$z_\mathrm{s} = 3.773$ (\S4). A flat FRW metric 
with $\Omega_\mathrm{m}=0.3$, $\Omega_{\Lambda}=0.7$ and 
$H_0 = 70 $\hubble~\kmsmpc is assumed.

%\begin{figure}[!t]
\begin{figure}
\psfig{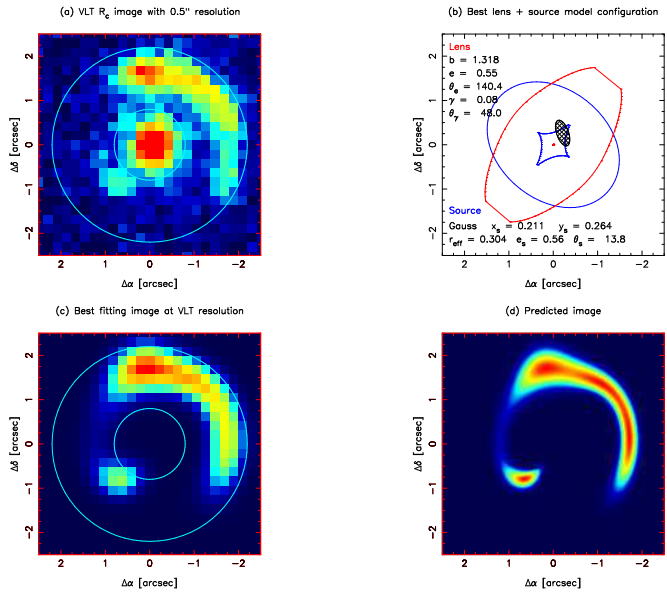}
\caption{(a) Image of the ring with 0\farcsec5 seeing, (b) best configuration
for the lens and the source, with the position of the caustics (cf. text for parameter definitions), (c) 
best-fitting
image convolved with the seeing, and (d) the predicted image at the resolution
of HST/ACS.}
\label{figmodel}
\end{figure}

\section{Modeling the gravitational lensing system}

Full details of the observational photometric and spectroscopic 
setups will be reported elsewhere (Cabanac et al. 2005).
Given the excellent image quality in the $R_\mathrm{c}$ band, with 0\farcs5 seeing, 
a detailed modeling of the lensing system can be carried out. We followed the
procedure pioneered by \citet{kochanek01} with two important
differences. First, not only we use
the detailed shape of the ring (ridge) and the flux along the ridge, but 
also the width of the 
ring along each radial direction. These three sets of observables
were incorporated into a likelihood function. Second, since 
we do not want to impose any {\sl a priori} information, the problem has a high
dimensionality: 5 parameters to describe the properties of the lens (given a 
mass profile: the normalisation, ellipticity and orientation, 
plus an external shear and its  
orientation) and 5 for the source (position with respect to the lens, 
effective radius, ellipticity and orientation). Hence, to guarantee that the 
maximum likelihood of the fit was found in the 10-dimensional parameter 
space, we coupled a state-of-the-art genetic algorithm developed by one of us 
with the {\tt gravlens} code \citep{keeton01}. Briefly stated, 
the parameters are coded in the chromosome of an individual which can produce 
two offsprings from its coupling with another individual \citep{charbonneau95}.
For each individual in each generation (i.e. one configuration in parameter 
space), we compute the resulting image, convolve it with the seeing 
of the VLT image, and measure its likelihood.
These steps are illustrated in Fig.~\ref{figmodel}. 
After a few thousand generations the genetic algorithm finds 
the absolute maximum of the likelihood, and several tests were made to assess
 that different runs always converged to the same solution. 
From the ensemble of generations and individuals 
the likelihood contours can be computed and the partial correlations and
degeneracies between parameters explored. The procedure will be presented 
in detail elsewhere.

Following \citet{kassiola93,kormann94,keeton98b}, a projected surface mass 
density for an isothermal ellipsoid (SIE) written in polar coordinates 
($r,\phi$) and expressed in units of the critical density for lensing,
\begin{eqnarray}
\kappa(r,\phi)&=&\frac{b}{2r}\left[\frac{1+q^2}{(1+q^2)-(1-q^2)cos2\phi)}\right],\nonumber
\end{eqnarray}
where $q\le1$ is the axial ratio. The radius $r$ and
the parameter $b$ both have dimensions of length (expressed here in arcsec).
$b$ is equal to the Einstein radius $\thetae$ in the spherical case, 
Following \citet{huterer04} we use an Einstein radius 

$\thetae = b \exp[(0.89e)^3]$, 

for a non-spherical SIE of ellipticity $e=1-q$.

Figure~\ref{figmodel} shows the best-fit example with $b=1\farcsec318\pm0.02$ 
and $e=0.55\pm 0.01$, yielding $\thetae = 1\farcsec48\pm 0\farcsec02$, that is,
11.8 \h70m1 kpc, with possibly a systematic error of 9\% due to the 
different possible normalisations of isothermal ellipsoids \citep{huterer04}.
Given the redshifts of the lens (\S3) and the source (\S4), the 
total mass within the Einstein radius is 
$\masse =(8.3 \pm 0.2) \, 10^{11}$\,\h70m1\,\msun. 
The lens is 20\% less massive than the elliptical
lens in MG 2016+112 at $z_\mathrm{l}=1.004$ \citep{koopmans02}.
Quite independently of the assumed elliptical potential, the derived
magnification of the source is 12.9. 
The mass ellipticity appears to be larger than the light ellipticity of 0.2
(\S3) but this is often found in other systems \citep{keeton98b}. 

%\begin{table}[!h]
\begin{table}
\caption{Photometric properties of the lens galaxy}
\begin{tabular}{lcccc}
\hline
Band & \multicolumn{2}{c}{effective radius $R_\mathrm{e}$}& $<\mu>^1$ &  $m_{\mathrm{tot}}$\\  
     & [\arcsec]& [\h70m1 kpc]  &    [mag$\cdot$arcsec$^{-2}$] & [mag]\\
\hline
$R_\mathrm{c}$&  0.36$\pm$0.05 & 2.9$\pm$0.4 &  22.20$\pm$0.10   & 22.42   \\
$i$ &   0.36\,~~~~~~~~~~ & 2.9\,~~~~~~~ &  20.70$\pm$0.20   & 20.92   \\
$J$ &  0.41$\pm$0.01 & 3.3$\pm$0.1  &  18.64$\pm$0.03   & 18.58 \\
$H$ &  0.43$\pm$0.01 & 3.4$\pm$0.1  &  17.70$\pm$0.03   & 17.54\\
$K_\mathrm{s}$ & 0.50$\pm$0.01 & 4.0$\pm$0.1 &  17.13$\pm$0.03   & 16.63\\
\hline
\end{tabular}\\
{$^1$\footnotesize $<\mu>$ is the average surface brightness, not to be confused
with the surface brightness at the effective radius $\mu(R_\mathrm{e})=\mu_{\mathrm{e}}$.}
\end{table}

The large number of constraints provided by the ring lifts the degeneracy
between the flattening of the potential and the external shear. For a wide
range of potentials we consistently find high flattenings and small shears.
The small value of the shear (Fig. 1b) $\gamma = 0.08 \pm 0.01$ is typical 
of those
found in loose groups of galaxies. Given its direction (48\deg) it is 
possible that the shear is caused by a faint nearby galaxy which 
lies $\zeta=5\farcsec7$ away at a position angle  $38\deg\pm0\fdeg5$ 
and with a photometric redshift of $\sim$1.27. If this is the case, 
the galaxy would have a SIS Einstein radius of \thetae = 2$\gamma \zeta$ = 0\farcsec91 and 
a projected velocity dispersion of 260\kms. It would imply a massive
galaxy, inconsistent with its observed brightness. On the other hand, 
the $R_\mathrm{c}$ and $B$ images show a diffuse population of blue galaxies
within one arcmin of the Einstein ring, and the near-IR images in 
$J$, $H$, and $K_\mathrm{s}$  provide evidence for a group of galaxies possibly 
associated with the lens or else with a proto-cluster at redshift 2 
as the photometric redshifts with {\tt HYPERZ} \citep{bolzonella00} 
show a bimodal probability distribution 
function with peaks at both redshifts.
There is no evidence, however, for a red sequence at $z \approx $ 1. 

\section{The lens galaxy}

\begin{figure}
\psfig{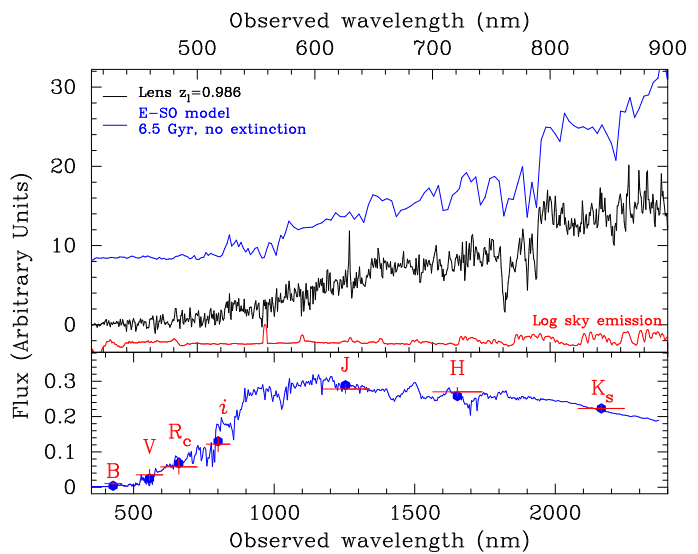}
\caption{Top : The spectrum of the lens (middle black line),
the underlying sky emission (bottom red line), and the 
{\tt HYPERZ}-derived best-fit synthetic spectrum (top blue line).
Bottom : Observed $BVR_\mathrm{c}iJHK_\mathrm{s}$ photometry and the 
spectral energy distribution of the best-fitting spectrum shown above. 
}
\label{figlens}
\end{figure}

The $BVR{_\mathrm{c}}iJHK_\mathrm{s}$  multicolour images obtained at ESO/VLT/La Silla 
and the optical spectrum of the lens 
(Figure~\ref{figlens}) constrain the spectral energy distribution  (SED)
unambiguously and allow the precise identification of the lens as a 
quiescent galaxy at redshift $z_\mathrm{l}=0.986\pm0.005$.\\
We used {\tt GALFIT} \citep{peng02} to fit the profile
of the lens, using a PSF-corrected profile with De Vaucouleur, Sersic 
and exponential parameterizations, masking the ring and using a range 
of sky level (average$\pm3\sigma$). This yields best-fit de 
Vaucouleur\footnote{Sersic profiles also fit the surface brightness with 
values of $n>3$, the dominant parameter in the Chi-square fit is the 
background subtraction, but since we find no hint of a disk in the $R$ band 
(blue restframe), we used de Vaucouleur profile parameterization} 
profiles with effective radii and average surface brightnesses given Table~1, 
and consistently small ellipticities $e=1-q=0.20\pm0.1$ in all bands. 
This distant lens appears rounder than the nearby ones, where the mean 
observed light ellipticity is 0.31$\pm$0.18 \citep{jorgensen95}.\\
%\begin{figure}[!t]
The fitted SED (Fig.~\ref{figlens}) allows one to infer properly the $k$ 
corrections following \citet{poggianti97}: we find 
$B_{\mathrm{rest}}-R_\mathrm{c \mathrm{obs}} = -0.65$, 
$B_{\mathrm{rest}}-i_{\mathrm{obs}} = 0.73$ and
$B_{\mathrm{rest}}-J_{\mathrm{obs}} = 3.08$, yielding a consistent 
$B_{\mathrm{rest}}= 21.7 \pm 0.1$.  Using other templates which
provide similar good fits to the SED yield the same result.
For sake of completeness, it should be noted that {\tt HYPERZ} 
can also fit (with a lower probability) bluer templates of {\em younger}
ages with extinctions of up to $A_\mathrm{V}=0.8$\,mag, assuming a \citeauthor{seaton79}
extinction law (1979). The definitive internal extinction of the lens can 
only be better constrained with a higher quality spectrum of higher 
resolution.\\
If we assume that the lens is an elliptical galaxy, we can compare its
measured optical properties with the nearby and high-redshift sample.
The rest-frame $M_\mathrm{B}=-22.3$ is twice brighter than the typical galaxies 
at $z \sim 1$ which have $M_\mathrm{B}^{\star} = -21.5 \pm 0.2$, and its rest-frame 
$U-V$ colour is redder (1.31) than the average $L_{\star}$ galaxy at this 
redshift ($U-V = 1.10 \pm 0.1$, \citealt{bell04}), while its rest-frame 
$U-B=0.18$ appears bluer than similar field ellipticals \citep{gebhardt03}.\\ 
The rest-frame absolute $B$-band surface brightness SB$_\mathrm{B}$, inferred from the 
$R_\mathrm{c}$ (resp. $J$) images, yield SB$_\mathrm{B}= 18.57\pm0.10$ (18.75) mag arcsec$^{-2}$ 
after the $(1+z)^4$ cosmological corrections.
The fundamental plane relation of elliptical galaxies can be written as 

$\Gamma_\mathrm{i} = \log (R_\mathrm{e}/\mathrm{kpc}) - \alpha \log \sigma - \beta\,\mathrm{SB}_\mathrm{B},$

where $R_\mathrm{e}$ is the effective radius, $\sigma$ is the velocity dispersion,
and we adopt $\alpha=1.25$, $\beta=0.32$ and $\Gamma_\mathrm{0} = -9.04 - \log $ 
\hubble\,   (Treu \& Koopmans \cite{treu04}). Taking the  velocity dispersion 
inferred from the lens modeling (\S 2) as $\sigma = 302 \pm 12$(syst) \kms, 
the evolution in the mass to light ratio becomes 

$\Delta \log(M/L_\mathrm{B}) = - (\Gamma_\mathrm{i} - \Gamma_\mathrm{0})/2.5\beta = -0.57 \pm 0.04$

where the uncertainty takes into account the systematics in the dispersion
velocity and the effective radius.  This evolution rate is consistent
with what is observed in field or cluster ellipticals  
\citep{gebhardt03,rusin03,koopmans03}, and within the scatter seen 
for massive galaxies \citep{treu05} at these redshifts.
It also implies a relatively large formation redshift, in a simplified
scenario where star formation proceeds in a single burst. This is
consistent with the SED-inferred stellar population. 
We note that the radio ring MG 1131+0456 (whose lens is at $z_\mathrm{l}$=0.844, 
\citealt{tonry00}) also yields a smaller evolutionary rate \citep{rusin03}.

Assuming that the evolution of the stellar mass-to-light ratio is the same as 
the effective one derived above, and taking a value
of $7.8\pm 2.2$ \hubble \mlbsun~ for the local galaxies, results in
 $(M_*/L_\mathrm{B}) = 2.1 \pm 0.1$ \hubble \mlbsun, entirely consistent
with relatively old stellar populations (as corroborated by the
 best fitting SED, Fig.~\ref{figlens}). 
% It implies that, within the effective radius of the lens,  
%there is no need for dark matter. 
However, at Einstein radius $\thetae$ the ratio 
$M(<\thetae)/L_\mathrm{B}(<\thetae) = 7.1\pm0.5$ \hubble \mlbsun.
%does require the presence of dark matter.  
%The reddish colours, plausible hint of the presence of dust,
%would decrease even further the $M/L$ ratio, and would
%increase the evolution rate (in prep. Cabanac et al. 2005).
% For a maximum $A_\mathrm{V}=0.6$, assuming $R_\mathrm{V}=3.1$,
% $A_\mathrm{B}=0.79$ and the evolution rate increases to  
% $\Delta \log(M/L_\mathrm{B}) = -0.92$.
% The systematic error can be reduced with higher resolution 
% spectra where the degeneracy between metallicity and reddening can be
% partly lifted.
% {\bf More about the photometric properties of the elliptical galaxy : mass
% in stars, dust vs Z,  J-K vs K (Holden 04), 
% I-Ks vs z (Glazebrook Abraham 04)}
% M_\mathrm{B}sun=5.50 (M_\mathrm{V}sun=5.5 (B-V)sun=0.65 -> M_\mathrm{B}=-21.65 => L_\mathrm{B}=6.1e10 L_\mathrm{B}sun

\section{The source galaxy}

\begin{figure}
\psfig{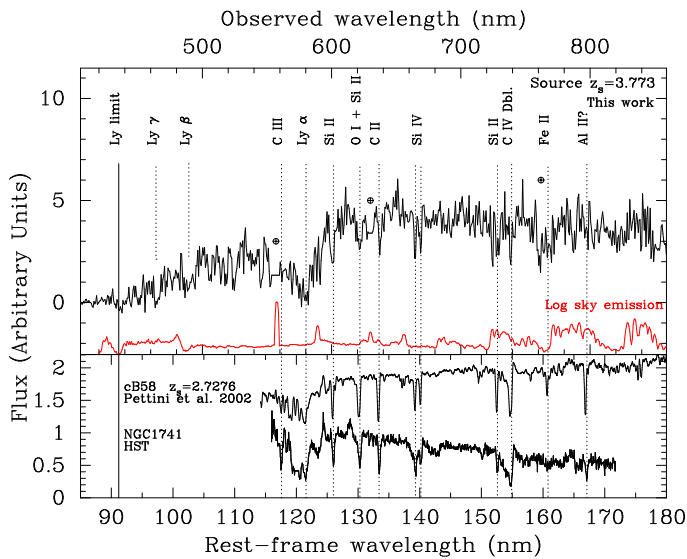}
\caption{Top : The spectrum of the source galaxy, showing strong lines
from the interstellar medium, yields a redshift of 3.773$\pm$0.003. 
Bottom: The shape of the continuum and the depth of the lines are 
consistent with a population of stars having past a major episode of 
star formation, very similar to the post-starburst galaxies shown
in comparison, both distant (cB58) or nearby (NGC 1741). The sky
emission and some telluric lines are also indicated.
}
\label{figsource}
\end{figure}

The VLT/FORS low-resolution spectrum of the lensed source (Figure~\ref{figsource})  
reveals a galaxy at a redshift 
$z_\mathrm{s} = 3.773\pm0.003$. Its apparent total 
magnitude $R_\mathrm{c}=22.6$ makes it about one magnitude brighter than the brightest 
Lyman-break galaxies (LBGs) found at $z \sim 3.5$ \citep{steidel03}. 
The source is sufficiently bright to make future follow-up kinematic and 
abundance studies of its interstellar medium  similar to what was possible on 
the $z_\mathrm{s}=2.73$ lensed galaxy cB58 \citep{pettini02}.
%\begin{figure}[!b]

The overall spectrum compares well with the high luminosity Lyman-break
galaxies \citep{shapley03} which have \lya~ in absorption at 
$z \sim 3$ or even with SDSS J1147$-$0250 \citep{bentz04} at a lower redshift. 
The continuum appears to be rather flat, bluer than cB58 but redder 
than NGC 1741, and it is remarkable that there are no emission lines 
(with the possible exception of C~{\sc iii}] $\lambda\lambda$ 190.7,190.9). 
It is clearly not like the H~{\sc ii} galaxy detected at a similar redshift 
($z=3.357$) in the Lynx arc \citep{fosbury03}. The flat continuum and the 
absence of \lya\ in emission, as seen in many nearby starbursts and about half 
the LBGs, is consistent with a post-burst stellar population, where the 
absorption stellar \lya~ line becomes important after a few million years 
\citep{vallsgabaud93} independently of metallicity or extinction.\\ 
Although the resolution of the discovery spectrum is not high enough 
to make a kinematic study of the different
lines, the signal-to-noise is sufficient to allow the identification of many 
photospheric and interstellar medium absorption features with the following 
rest-frame equivalent widths (in nm) : 
C~{\sc ii} $\lambda$133.5 (0.15$\pm$0.02),       
C~{\sc iv} $\lambda\lambda$154.8,155.1 (0.08$\pm$0.02),         
O~{\sc i}/Si~{\sc ii} [blend]$\lambda\lambda$ 130.2,130.4 (0.17$\pm$0.02), 
Si~{\sc ii}  $\lambda$126.0  (0.30$\pm$0.04), 
Si~{\sc ii}  $\lambda$152.7  (0.25$\pm$0.04), 
Si~{\sc iv}  $\lambda$139.4  (0.13$\pm$0.02), 
Si~{\sc iv}  $\lambda$140.3  (0.08$\pm$0.02).  
There are also tentative detections of other important lines, such as 
Fe~{\sc ii} $\lambda$160.8, C~{\sc iii} $\lambda$117.6, and 
Al~{\sc ii} $\lambda$167.0, along with possible features produced by 
P~{\sc v} and O~{\sc i}.
The metallicity index at 143.5nm is contaminated by sky lines, but 
the pattern that emerges from these values,  although preliminary, 
is that the gas-phase abundances of these interstellar absorption lines are 
rather high, especially in comparison with cB58 \citep{pettini02}. 
In this respect, the source galaxy appears  more similar to the LBGs 
at $z \sim 5$ \citep{ando04} 
where the ISM lines are stronger than at $z=3$ (with the notable 
exception of carbon). These indications point to a very rapid enrichment by
type II supernovae, associated to bursts of star formation, in many
respects similar to the pattern seen in star-forming galaxies at $z \sim 2.2$ 
(Shapley et al \cite{shapley04}).

The flux at 150\,nm has been traditionally used to measure the star formation
rate from UV spectra. Calibrating on the well-measured $R_\mathrm{c}$ magnitude, 
we derive a flux density at 150\,nm of 2.15\,$\mu$Jy (assuming a conversion 
factor of 2875\,Jy for a $R_\mathrm{c}=0$ source), which translates into 
$\mathrm{SFR}_{\mathrm{UV}} \approx 31\,(A/12.9)^{-1}$ \hm2 \msun\, a$^{-1}$, 
adopting the standard relation (\citealt{kennicutt98}, 
and where $A$ is the gravitational amplification 
produced by the lens (\S2). There are two important caveats. First,  this 
could be a lower value because we have neglected the dust absorption, both 
internal to the source galaxy and in the lens galaxy. Second, 
the relation only holds for a constant star formation rate, while it seems
more likely that star formation proceeded in a series of bursts. Catching the
galaxy after the burst has finished, or is decreasing, implies that the UV
luminosity is a strong function of age and cannot be properly related to the 
rate that produced these stars unless their age can be measured.  
All in all, the large inferred rate is better used for comparison with other 
galaxies, in which case this source appears to have been extremely active. 
Although it is more than an order of magnitude below the extremely bright 
SDSS J1147$-$0250 \citep{bentz04} it is entirely consistent with the rates 
inferred for LBGs at $z \sim 3$ \citep{shapley03} or the 
$\sim 40 $ \msun\, a$^{-1}$ rate of cB58 \citep{pettini02},
 and would explain the rapid metal 
enrichment of its interstellar medium sketched above.
  
The modeling of the lensing system (\S1) also yields constraints on the
effective size of the source galaxy, $\theta_{\mathrm{eff}}= 0\farcsec304$ 
(2.16\,\h70m1\,kpc), assuming an elliptical gaussian source\footnote{the 
actual shape and position angle of the source are not strongly constrained
by present images} . Other elliptical shapes yield similar values. 
The compactness of the region which gives rise to this emission, combined 
with its tentative high metallicity may perhaps be associated with a 
progenitor of a present-day bulge.

Additional observations are clearly required to further constrain this lens, 
and to study the interstellar medium and the stellar populations  
at a look-back time of 88\% of the present age of the universe.
 
\begin{acknowledgements}
We are very grateful to C. Keeton for a copy of his {\tt gravlens} code 
and to the anonymous referee for a detailed verification of our calculations
which substantially improved the paper.
\end{acknowledgements}
\bibliographystyle{aa}
\bibliography{lens}
\end{document}